\documentclass[letterpaper, 10 pt, conference]{ieeeconf}  

\IEEEoverridecommandlockouts                              

\usepackage{graphics} 
\usepackage{epsfig} 
\usepackage{amsmath} 
\usepackage{amssymb}  
\usepackage{enumerate}
\usepackage{svg}
\usepackage{caption}

\usepackage{graphicx}
\usepackage{float} 
\usepackage{subfigure}

\usepackage{booktabs} 
\usepackage{multirow} 
\usepackage{array}

\title{\LARGE \bf

SEAL: Safety Enhanced Trajectory Planning and Control Framework for Quadrotor Flight in Complex Environments
}

\author{Yiming Wang$^{\dagger}$, Jianbin Ma$^{\dagger}$, Junda Wu, Huizhe Li, Zhexuan Zhou, Youmin Gong, Jie Mei$^{*}$ and Guangfu Ma
\thanks{$^{\dagger}$These two authors contribute equally.}
\thanks{* Corresponding Author.}
\thanks{Yiming Wang, Jianbin Ma, Junda Wu, Huizhe Li, Zhexuan Zhou, Youmin Gong, Jie Mei and Guangfu Ma are with the School of Intelligence Science and Engineering, Harbin Institute of Technology, Shenzhen, China. For correspondence:{\tt\small jmei@hit.edu.cn}}
}

\begin{document}

\captionsetup[figure]{labelfont={bf},labelformat={default},labelsep=period,name={Fig.}}

\maketitle
\thispagestyle{empty}
\pagestyle{empty}

\begin{abstract}

For quadrotors, achieving safe and autonomous flight in complex environments with wind disturbances and dynamic obstacles still faces significant challenges. Most existing methods address wind disturbances in either trajectory planning or control, which may lead to hazardous situations during flight. The emergence of dynamic obstacles would further worsen the situation. 
Therefore, we propose an efficient and reliable framework for quadrotors that incorporates wind disturbance estimations during both the planning and control phases via a generalized proportional integral observer. 
First, we develop a real-time adaptive spatial-temporal trajectory planner that utilizes Hamilton-Jacobi (HJ) reachability analysis for error dynamics resulting from wind disturbances. By considering the forward reachability sets propagation on an Euclidean Signed Distance Field (ESDF) map, safety is guaranteed.
Additionally, a Nonlinear Model Predictive Control (NMPC) controller considering wind disturbance compensation is implemented for robust trajectory tracking. Simulation and real-world experiments verify the effectiveness of our framework.
The video and supplementary material will be available at https://github.com/Ma29-HIT/SEAL/.


\end{abstract}




\section{Introduction}

Recently, autonomous quadrotors have been deployed in various tasks, such as exploration \cite{tang2023bubble}, target tracking \cite{ji2022elastic}, and aerial delivery \cite{li2023autotrans}. In such situations, quadrotors will inevitably encounter wind disturbances, which pose critical risks to safe flight. 
The emergence of dynamic obstacles further exacerbates the challenges. 

\begin{figure}[htbp]
\centering
\includegraphics[width=0.45\textwidth]{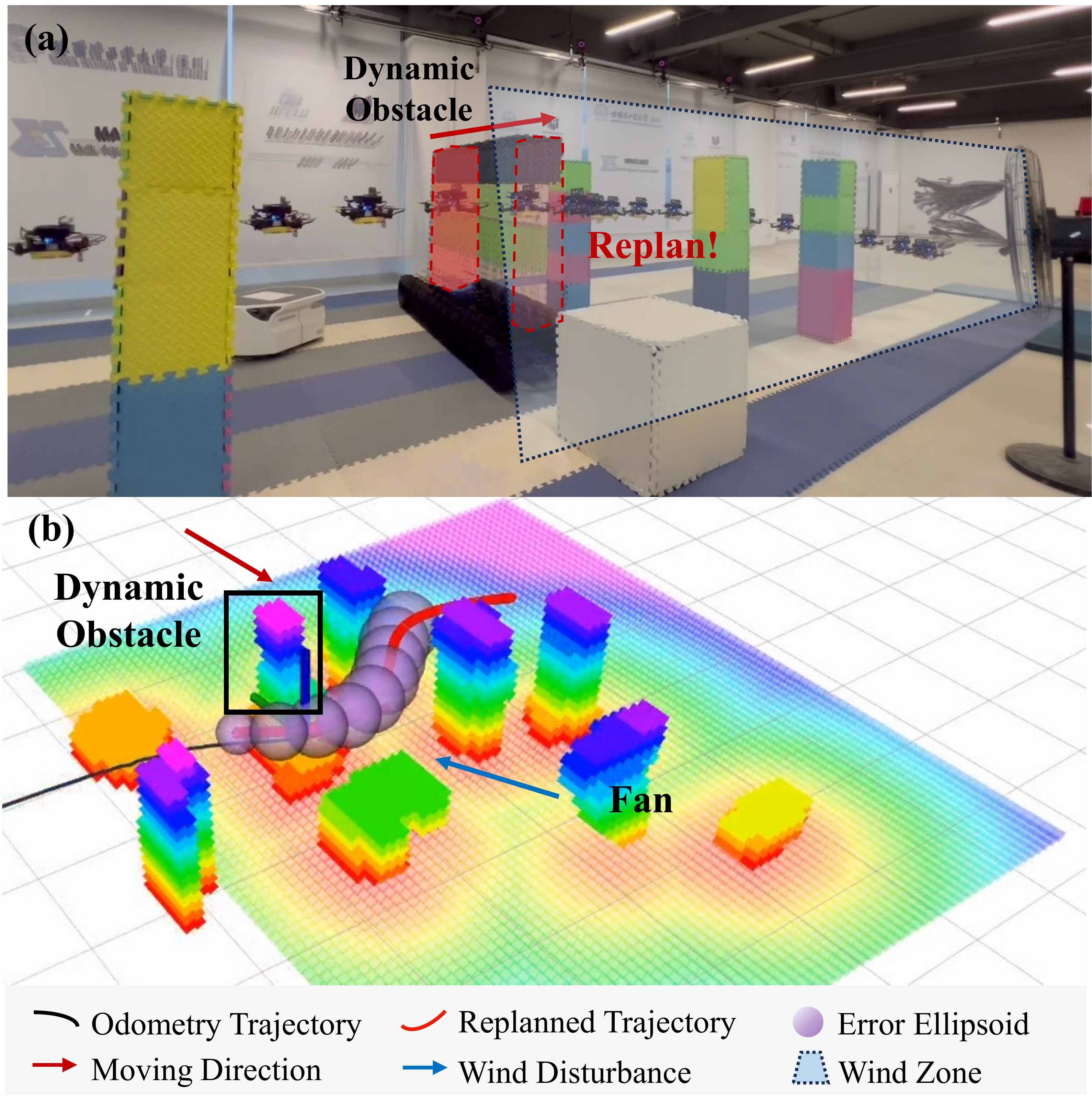}
\caption{Snapshot of an indoor real flight test. (a) The dynamic obstacle (in red dashed line) is moving towards the quadrotor, while the quadrotor enters the wind zone that pushes it towards the obstacles. Our planner replans a safe and feasible trajectory (in red line). (b) The visualization of the flight.}
\label{flight_snap}
\end{figure}

Quadrotor planning methods \cite{yang2022far,lu2024fapp} have demonstrated impressive performance in autonomous flight. However, wind disturbances can induce trajectory tracking errors, which present significant challenges for these planning methods. 
Hence, it is necessary to plan a safe and feasible trajectory and ensure accurate tracking in environments affected by wind disturbances. 
Recent advances in safety-aware planning 
\cite{wu2021external}, \cite{liu2023tight}, \cite{NarrowSpaces} and robust control \cite{xu2024fixed}, \cite{li2023autotrans} have improved anti-disturbance capabilities of quadrotors. However, treating planning and control as distinct entities could pose potential risks when confronted with challenges in complex environments.
For the planner, our primary requirement is the ability to ensure safe flight in windy conditions and dynamic environments. Additionally, the planner must be lightweight enough for real-time operation. Current methods 
	\cite{wu2021external,NarrowSpaces} demand the utilization of flight corridors to guarantee safety during the flight. Consequently, the quadrotor's movement is confined within the corridor, which is restrictive and unsuitable for avoiding dynamic obstacles. Even though a specific approach \cite{liu2023tight} manages to avoid dynamic obstacles while considering the corridors, the quadrotor must navigate within the cramped space of the corridor to avoid the obstacles, leading to a compromise in flight flexibility.  

As for robust trajectory tracking control, advanced controllers such as 
\cite{xu2024fixed, torrente2021data, hanover2021performance} have shown remarkable results.
However, relying solely on controller design can be detrimental to ensuring the safe flight of quadrotors since they cannot proactively replan the trajectory when persistent disturbances exceed their local compensation capacities. 
These limitations highlight the necessity for a systematic planning-control framework that addresses disturbances holistically.

In this work, we aim to propose \textbf{SEAL}, a \textbf{S}afety-\textbf{E}nhanced co-design trajectory pl\textbf{A}nning and contro\textbf{L} framework to handle those challenges systematically.  
In the planning pipeline, a kinodynamic A* algorithm is first employed to generate an initial collision-free path. 
We deploy the forward reachable set (FRS) to propagate the uncertainty of wind disturbances.
To facilitate the computation, we extend the FRS ellipsoidal approximation method \cite{seo2019robust} into a real-time spatial-temporal trajectory planner, which incorporates penalty terms for both static and dynamic obstacles as well as dynamic feasibility jointly. 
Moreover, a generalized proportional integral observer is developed to perform online estimation of wind disturbances. 
An NMPC controller is then implemented, and wind disturbance estimation is used to compensate for the NMPC prediction model.
The contribution of this work is summarized as follows
\begin{itemize}
\item [1)] We propose a disturbance-aware trajectory planning and control framework that enhances the safety of quadrotors during autonomous flight in environments with wind disturbances and dynamic obstacles.
\item [2)] We employ a real-time spatial-temporal planning method in a dynamic obstacle environment that considers the influence of wind disturbances through the FRS propagation. A disturbance observer-based NMPC controller is then implemented to enhance the system's robustness.
\item [3)] Simulations and real-world experiments are both performed to validate our method. We plan to release our code to the robotics society.
\end{itemize}

\section{Related Works}
Various planning and control methods have been proposed to ensure the safe and autonomous flight of quadrotors in environments characterized by wind disturbances and dynamic obstacles. For adaptive trajectory generation, Hamilton-Jacobi reachability analysis is used to predict the reachable set of the quadrotors facing unknown but bounded disturbances.
Seo et al. \cite{seo2019robust} approximate the error state FRS using ellipsoid approximation, which facilitates real-time computations. In \cite{wang2021estimation}, a trajectory generation method utilizes B-splines and considers the control error bound imposed by ego airflow disturbances. However, constructing the airflow disturbance field model depends on the pre-collection of flight data.

Model predictive control (MPC) methods have been extensively studied as a planning approach, where the incorporation of flight corridors enables MPC to achieve robust trajectory planning and safe flight \cite{wu2021external, liu2023tight, NarrowSpaces}. 
PE-Planner \cite{qiu2024pe} constructs control barrier functions for both static and dynamic obstacles, imposing these constraints to model predictive contouring control and achieving average speeds of up to $6.98 m/s$. Simultaneously, the disturbance observer is utilized to compensate for the dynamics model of quadrotors. IPC \cite{liu2023integrated} is an MPC-based integrated planning and control framework that operates at a high frequency of $100$ Hz in the presence of suddenly appearing objects and disturbances. Most of the referenced works require the construction of flight corridors, which leads to more conservative operation of quadrotors and hinders the full utilization of obstacle-free flight areas. Additionally, methods that do not rely on flight corridors necessitate obstacle avoidance constraints in the MPC constraint set, which limits the computation frequency and prevents quadrotors from flying autonomously in complex environments.

In terms of robust trajectory tracking controllers, geometric control \cite{lee2010geometric} is considered effective due to its simple and intuitive design process. Additionally, nonlinear tracking controllers such as sliding mode 
\cite{xu2006sliding} and backstepping 
\cite{madani2006control} have been proposed to attain improved performance. 
However, these methods cannot explicitly address constraints such as control input limitations. NMPC is widely used for trajectory tracking due to its ability to effectively manage the complex constraints and generate optimal control outputs. Li et al. \cite{li2023autotrans} developed an external force estimator by formulating a nonlinear least squares problem; the estimated force is then used to correct the dynamics model within the NMPC. Xu et al. \cite{xu2024fixed} developed a fixed-time disturbance observer to update the dynamic model of NMPC, leading to nearly a 70\% reduction in trajectory tracking error compared to standard NMPC.






\section{Preliminaries}
\begin{figure*}[htbp]
\centering
\includegraphics[width=1.0\textwidth]{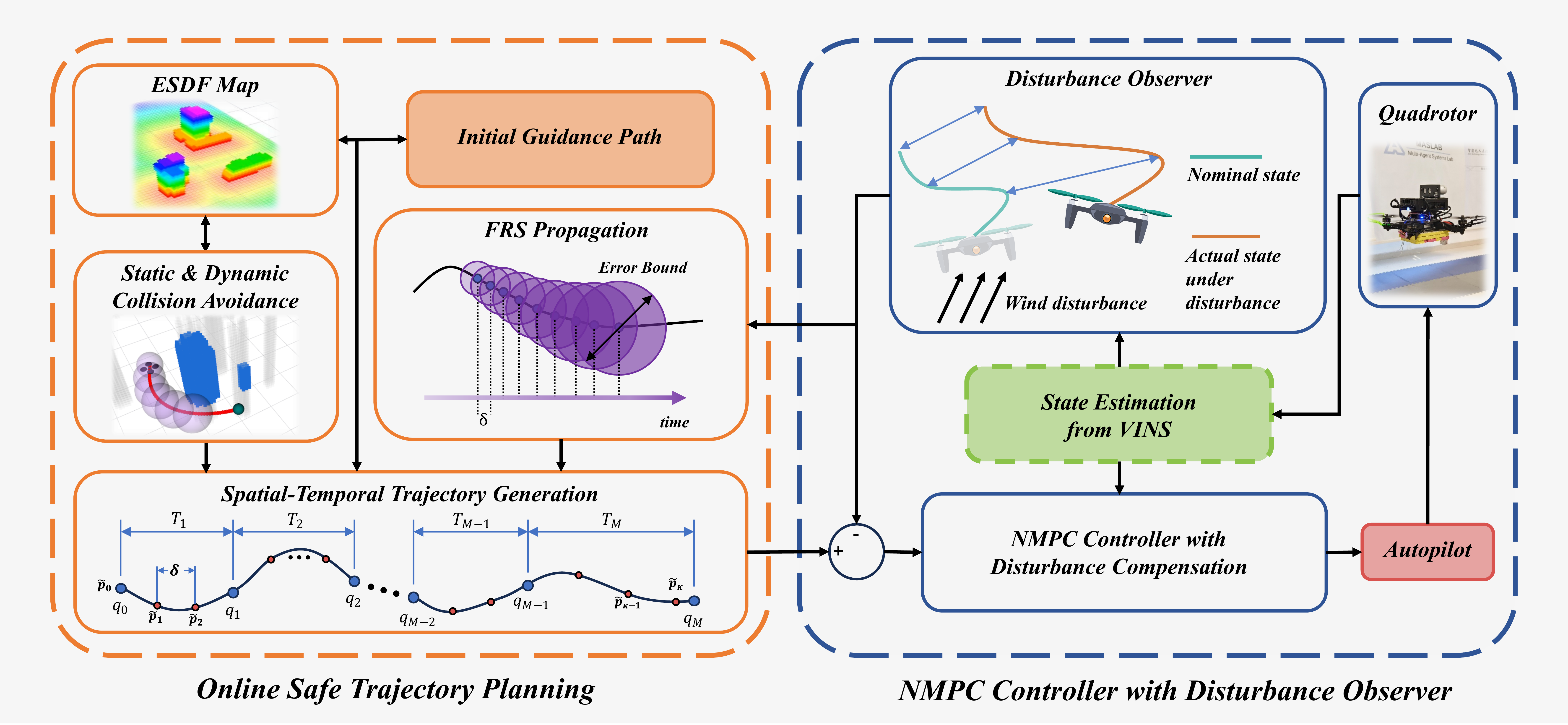}
\caption{The system overview diagram of SEAL. The disturbance observer provides an online estimation of wind forces while the planner and controller handle the disturbance jointly to generate a safe, smooth, and feasible trajectory and track it accurately.}
\label{fig:sys_overview}
\end{figure*}

\subsection{Quadrotor Dynamic Model}

We consider the six-degrees-of-freedom dynamic model of the quadrotor with Euler angles for attitude control. 
The quadrotor's state is $\boldsymbol{x}=[\boldsymbol{p}^\mathrm{T},\boldsymbol{v}^\mathrm{T},\phi,\theta,\psi]^\mathrm{T}\in\mathbb{R}^9$, where $\boldsymbol{p}\in\mathbb{R}^3$ and 
$\boldsymbol{v}\in\mathbb{R}^3$ represent the position and velocity, while $\phi,\theta,\psi\in\mathbb{R}$ denote the roll, pitch, and yaw angles.
The quadrotor's control input is $\boldsymbol{u}=[T_c,\dot{\phi}_c,\dot{\theta}_c,\dot{\psi}_c]^\mathrm{T}\in\mathbb{R}^4$, where $T_c\in\mathbb{R}$ denotes the collective thrust along the body frame, and $\dot{\phi}_c,\dot{\theta}_c,\dot{\psi}_c\in\mathbb{R}$ are the command rates of the Euler angles.
\par The nonlinear dynamic model of the quadrotor is given as
\begin{subequations}\label{eqn: non_lin_dynamics}
\begin{align}
\dot{\boldsymbol{p}} &= \boldsymbol{v},\\
\dot{\boldsymbol{v}} &= \frac{1}{m} \left( \boldsymbol{R} \begin{bmatrix} 0 \\ 0 \\ T_c \end{bmatrix} - \boldsymbol{F}_{\text{drag}} + \boldsymbol{F}_{\text{dist}} \right) - \boldsymbol{g},\\
\dot{\phi} &= \dot{\phi}_c, \\
\dot{\theta} &= \dot{\theta}_c, \\
\dot{\psi} &= \dot{\psi}_c,
\end{align}
\end{subequations}
where $m\in\mathbb{R}$ is the mass of the quadrotor, 
$\boldsymbol{g}=\left[0,0,9.81\right]^{\mathrm{T}}$ is the gravitational acceleration, 
$\boldsymbol{R}\in\mathbb{R}^{3\times 3}$ is the rotation matrix from the body frame to the world frame, $\boldsymbol{F}_{\text{drag}}\in\mathbb{R}^3$ is the drag force, and $\boldsymbol{F}_{\text{dist}}\in\mathbb{R}^3$ is the disturbance. The nonlinear dynamic model \eqref{eqn: non_lin_dynamics} is abbreviated as 
$\dot{\boldsymbol{x}} = f(\boldsymbol{x},\boldsymbol{u},\boldsymbol{F}_{\text{dist}})$.
To simplify the dynamics, the drag force is formulated as the following first-order drag model \cite{kai2017nonlinear}  
\begin{equation}
\boldsymbol{F}_{\text{drag}} = \boldsymbol{R} \boldsymbol{D}_{\text{drag}} \boldsymbol{R}^\mathrm{T} \boldsymbol{v},
\end{equation}
where $\boldsymbol{D}_{\text{drag}}\in\mathbb{R}^{3\times3}$ denotes the drag coefficient matrix.

\subsection{Parameterization and Propagation of FRSs}  

The forward reachable set (FRS) refers to the collection of states that a system can attain under all possible disturbances and control inputs \cite{seo2019robust}. By leveraging FRSs, the possible flight region that a quadrotor may reach under the disturbances is evaluated. 
Keeping this region outside the obstacles enables the generation of safer flight trajectories.

To simplify the calculation, a linearized system is first constructed, and the error dynamics are derived.
Denote the nominal states, inputs, and external force as $\underline{\boldsymbol{x}}(t)$, $\underline{\boldsymbol{u}}(t)$, $\underline{\boldsymbol{F}}_{\mathrm{dist}}$, respectively. Then, the error state is $\boldsymbol{e}(t)\triangleq \boldsymbol{x}(t)-\underline{\boldsymbol{x}}(t)$. The linearization error is assumed to be negligible and the feedback control policy is adopted as $\boldsymbol{u}(t) = \boldsymbol{K}(t)\boldsymbol{e}(t)+\underline{\boldsymbol{u}}(t)$, where $\boldsymbol{K}(t)$ is the feedback gain.
The error dynamics can thus be expressed as
\begin{equation}\label{sys:error_dynamics}
\boldsymbol{\dot{e}}(t)=\boldsymbol{\Phi}(t)\boldsymbol{e}(t)+\boldsymbol{D}(t)\boldsymbol{F}_{\mathrm{dist}}(t),
\end{equation}
where $\boldsymbol{\Phi}(t)=\boldsymbol{A}(t)+\boldsymbol{B}(t)\boldsymbol{K}(t)$, 
$\boldsymbol{A}(t)\triangleq\partial f/\partial\boldsymbol{x}|_{(\underline{\boldsymbol{x}},\underline{\boldsymbol{u}},\underline{\boldsymbol{F}}_{\mathrm{dist}})}$, 
$\boldsymbol{B}(t)\triangleq\partial f/\partial\boldsymbol{u}|_{(\underline{\boldsymbol{x}},\underline{\boldsymbol{u}},\underline{\boldsymbol{F}}_{\mathrm{dist}})}$, 
$\boldsymbol{D}(t)\triangleq\partial f/\partial\boldsymbol{F}_{\mathrm{dist}}|_{(\underline{\boldsymbol{x}},\underline{\boldsymbol{u}},\underline{\boldsymbol{F}}_{\mathrm{dist}})}$.

According to \cite{seo2019robust}, with Hamilton-Jacobi (HJ) reachability analysis, the analytic solution of error state FRS 
can be derived, which is further approximated and propagated as ellipsoidal bounds.
Specifically, the system equation is discretized with a sampling time $\delta$ over $N$ time steps. Define $\boldsymbol{\Phi}^k = \boldsymbol{\Phi}(k\delta)$, $\boldsymbol{D}^k = \boldsymbol{D}(k\delta)$, 
and denote the shape matrix of ellipsoidal approximation 
as $\boldsymbol{Q}_e^k$, where time stamp $k\in\{0,1,\dots,N-1\}$. 
The initial error state set is defined as an ellipsoidal set whose shape matrix is $\boldsymbol{Q}_{\mathrm{0}}^0$. 
To propagate the disturbances along the trajectory, $\boldsymbol{Q}_d^k$ is introduced to denote the shape matrix of FRS caused by the disturbances, which is given by
\begin{equation}\label{eqn:Q_d}
\boldsymbol{Q}_d^k=\left(\sum_{i=1}^{3}\sqrt{tr(\boldsymbol{Q}_i^k)}\right)\left(\sum_{i=1}^{3}\frac{\boldsymbol{Q}_i^k}{\sqrt{tr(\boldsymbol{Q}_i^k)}}\right),
\end{equation}
where $\boldsymbol{Q}_i^k$ represents the ellipsoidal approximation of the FRS resulting from the $i$\textsuperscript{th} channel of the disturbance, and it is the solution of the following Lyapunov equation:
\begin{equation}\label{lyapunov}
    \begin{gathered}
        -\boldsymbol{\Phi}^k\left(\boldsymbol{Q}_i^k-\varepsilon \delta^{2}I\right)-\left(\boldsymbol{Q}_i^k-\varepsilon \delta^{2}I\right)\boldsymbol{\Phi}^{k, \mathrm{T}}\\
        =\exp(-\boldsymbol{\Phi}^k\delta)\boldsymbol{N}_{i}^k\exp(-\boldsymbol{\Phi}^{k, \mathrm{T}}\delta)-\boldsymbol{N}_{i}^k,
\end{gathered}
\end{equation}
where $\boldsymbol{N}_{i}^k\triangleq \delta\bar{b}_{i}^{2}\boldsymbol{D}_{i}^k\boldsymbol{D}_{i}^{k, \mathrm{T}}$, $\varepsilon$ is a positive scalar indicating the conservativeness, $\bar{b}_i$ denotes the upper bound of the $i$\textsuperscript{th} channel of the disturbance, and $\boldsymbol{D}_{i}^k$ denotes the $i$\textsuperscript{th} column of $\boldsymbol{D}^k$.

With \eqref{eqn:Q_d}, the propagation law for the initial shape matrix can be derived as
\begin{equation}\label{eqn:propagate_of_Q_0}
    \boldsymbol{Q}_0^k =\boldsymbol{Q}_{0}^{k-1}\oplus \boldsymbol{Q}_{d}^{k-1},
\end{equation}
where the Minkowski sum $\oplus$ for shape matrices of two ellipsoids centered on the same point is defined as 
\begin{equation}
    \boldsymbol{Q}_1\oplus\boldsymbol{Q}_2=\bigl(1+\tfrac{b}{a}\bigr)\boldsymbol{Q}_1+\bigl(1+\tfrac{a}{b}\bigr)\boldsymbol{Q}_2,
\end{equation}
where $\boldsymbol{Q}_1,\boldsymbol{Q}_2$ are positive definite matrices, 
$a=\sqrt{tr(\boldsymbol{Q}_1)}$ and $b=\sqrt{tr(\boldsymbol{Q}_2)}$.

By combining \eqref{eqn:Q_d} and \eqref{eqn:propagate_of_Q_0}, the expression of $\boldsymbol{Q}_e^k$ is shown as
\begin{equation}\label{eqn:qdist}
    \boldsymbol{Q}_e^k=\exp(\boldsymbol{\Phi}^k\delta)(\boldsymbol{Q}_0^k\oplus \boldsymbol{Q}_d^k)\exp(\boldsymbol{\Phi}^{k, \mathrm{T}}\delta). 
\end{equation}
 
In our trajectory planning method, we consider only the uncertainty associated with the quadrotor's position. Therefore, we extract the first three rows and columns from $\boldsymbol{Q}_e^k$, forming a 
$3\times 3$ matrix $\boldsymbol{Q}_{\mathrm{dist}}^k$ for subsequent collision avoidance constraints design.


\subsection{System Overview}
The overall framework is depicted in Fig. \ref{fig:sys_overview}. A disturbance observer is introduced to estimate the wind force.
Then, the forward reachable sets (FRSs) are established, and a spatial-temporal trajectory optimization method that considers the propagated error bounds is employed to generate a smooth, safe, and dynamically feasible trajectory. Finally, a nonlinear model predictive controller (NMPC) is implemented to track the planning trajectory, and disturbance estimation is incorporated to update the prediction model.

\section{Real-Time Trajectory Planning with Wind Disturbances and Dynamic Environments}

\subsection{Trajectory Definition and Optimization Problem Formulation}

The differential flatness \cite{mellinger2011minimum} of the quadrotor enables us to define its motion as a piecewise polynomial trajectory.
Specifically, for a $3$-dimensional $M$-piece trajectory, the $i$\textsuperscript{th} piece $\boldsymbol{p}_i(t)$ is denoted by
\begin{equation}
  \boldsymbol{p}_i(t) = \mathbf{c}_i^\mathrm{T}\boldsymbol{{\beta}}(t),\quad\forall t \in [0,T_i],
\end{equation}
where $\boldsymbol{c}_i\in\mathbb{R}^{6\times 3}$ is the polynomial coefficient, $\boldsymbol{\beta}(t) = [t^0, t^1,...,t^5]^\mathrm{T}$ is the natural basis, and $T_i$ is the duration of this piece. Thus, the whole trajectory is expressed as
$\boldsymbol{p}(t) = \boldsymbol{p}_i(t-t_{i-1})$, where $i\in\{1,\dots,M\}$, $ t\in [t_{i-1},t_i]$, and $t_{i-1}$ is the end time of $\boldsymbol{p}_{i-1}(t)$.

The minimum control (\textbf{MINCO}) class \cite{wang2022geometrically} is subsequently adopted, which allows for the spatial-temporal parameter decoupling through the following linear-complexity mapping
\begin{equation}\label{mapping}
  \boldsymbol{c}=\mathcal{M}(\boldsymbol{q}, \boldsymbol{T}),
\end{equation}
where $\boldsymbol{c} = [\boldsymbol{c}_1^\mathrm{T},\dots,\boldsymbol{c}_M^\mathrm{T}]^\mathrm{T}\in\mathbb{R}^{6M\times 3}$ is the polynomial coefficients, $\boldsymbol{q}=\left(q_1, \cdots, q_{M-1}\right) \in \mathbb{R}^{3 \times(M-1)}$ is the intermediate points between pieces, $\boldsymbol{T}=\left(T_1, \cdots, T_M\right) \in \mathbb{R}_{>0}^M $ denotes the duration of each piece.  By \eqref{mapping},  any second-order continuous cost function $\mathcal{F}(\mathbf{c}, \mathbf{T})$ with available gradient can apply to MINCO. 
Thus, we can conduct the optimization over objective $\mathcal{J}$ with the gradients $\partial\mathcal{J}/\partial\mathbf{q}$ and $\partial\mathcal{J}/\partial\mathbf{T}$, which are propagated from $\partial\mathcal{F}/\partial\mathbf{c}$ and $\partial\mathcal{F}/\partial\mathbf{T}$.

For MINCO, we can formulate a nonlinear optimization problem that can be solved in real-time:
\begin{equation}
\min_{\boldsymbol{q}, \boldsymbol{T}} \sum_{*} \lambda^* \mathcal{J}^* + \int_{0}^{T_{\Sigma}} \| \boldsymbol{p}^{(3)}(t) \|^2 dt + \rho \cdot T_{\Sigma}.
\end{equation}
The first term includes the penalty cost $ \mathcal{J}^*$ and the corresponding weights $\lambda^*$ for the quadrotor's constraints. Superscripts $* = \{s,d,f\}$, where $s$
 denotes the static obstacle avoidance, $d$ denotes the dynamic obstacle avoidance, and
 $f$ denotes the dynamic feasibility. The second term is designed for improving smoothness of the trajectory, while the total time $T_\Sigma$ in the last term is utilized to exploit the trajectory's aggressiveness. An open source library L-BFGS\footnote{https://github.com/ZJU-FAST-Lab/LBFGS-Lite}
\cite{liu1989limited} is adopted to solve the optimization problem.
\subsection{Quadrotor's Constraints}
To ensure the effective computation of the constraint violations, we use constraint transcription, converting the continuous constraints into the ones sampling at finite points \cite{quan2023robust}. Since the propagation of error bounds is carried out at a fixed step size $\delta$, the sampling points are set at a fixed-time-interval fashion along the whole trajectory. 
The sampling number $\kappa$ of the constraint points is determined by $\kappa =  \lfloor T_{\Sigma}/\delta\rfloor$.
Given the $k$\textsuperscript{th} $(k=0,1,\dots,\kappa)$ constraint point $\overset{\circ}{\boldsymbol{p}}_k$, it can be located in the unique piece $\boldsymbol{p}_j(t)$ by
\begin{equation}
 \overset{\circ}{\boldsymbol{p}}_k = \boldsymbol{p}_j \left( k\delta - T_l \right), T_l\leq k\delta < T_l + T_j,
\end{equation}
where $T_l = \sum_{i=1}^{j-1} T_i$ is the total duration of the first $j-1$ pieces, $k\delta$ refers to the time relative to the start time of the entire polynomial trajectory, while $t=k \delta - T_l$ refers to the time relative to the start time of the $j$\textsuperscript{th} piece.
Then, the penalty cost is expressed as
\begin{align}
\mathcal{J}^*= \delta \sum_{k=0}^{\kappa} \omega_k \mathcal{P}^*(k \delta) + \frac{1}{2} (T_{\Sigma} - \kappa \delta) [ \mathcal{P}^*(\kappa \delta) + \mathcal{P}^*(T_{\Sigma}) ],
\end{align}
where $(\omega_0, \omega_1, \dots, \omega_{\kappa - 1}, \omega_{\kappa}) = ( 1/2, 1, \dots, 1, 1/2 )$ is the coeffient following the trapezoidal rule, $\mathcal{P}^*$ is the penalty of the constraints.
We define the set of all constraint points located within the $j$\textsuperscript{th} piece as $\mathcal{O}$, then, the gradient of $\mathcal{J}^*$ with respect to $\boldsymbol{c}_j$ and $T_i(i\leq j)$  can be expressed with the chain rule:
\begin{align}
\frac{\partial \mathcal{J}^*}{\partial \boldsymbol{c}_j} 
&=\sum_{k\in \mathcal{O}} \left(\frac{\partial \mathcal{J}^*}{\partial \mathcal{P}^*} \frac{\partial \mathcal{P}^*} 
 {\partial \psi^*} \frac{\partial \psi^*}{\partial \overset{\circ}{\boldsymbol{p}}_k}\frac{\partial \overset{\circ}{\boldsymbol{p}}_k}{\partial \boldsymbol{c}_j}\right)\nonumber\\
&= \sum_{k\in \mathcal{O}} \left.\left(\frac{\partial \mathcal{J}^*}{\partial \mathcal{P}^*} \frac{\partial \mathcal{P}^*} 
 {\partial \psi^*} \frac{\partial \psi^*}{\partial {\boldsymbol{p}}_j(t)}\frac{\partial \boldsymbol{p}_j(t)}{\partial \boldsymbol{c}_j}\right)\right\vert_{t = k\delta-T_l}, \\
\frac{\partial \mathcal{J}^*}{\partial T_i} 
&=\sum_{k\in \mathcal{O}} \left(\frac{\partial \mathcal{J}^*}{\partial \mathcal{P}^*} \frac{\partial \mathcal{P}^*}{\partial \psi^*} \frac{\partial \psi^*}{\partial \overset{\circ}{\boldsymbol{p}}_k} \frac{\partial \overset{\circ}{\boldsymbol{p}}_k}{\partial t}\frac{\partial t}{\partial T_i} \right) \nonumber\\
&= \sum_{k\in \mathcal{O}} \left.\left(\frac{\partial \mathcal{J}^*}{\partial \mathcal{P}^*} \frac{\partial \mathcal{P}^*}{\partial \psi^*} \frac{\partial \psi^*}{\partial {\boldsymbol{p}}_j(t)} \frac{\partial {\boldsymbol{p}}_j(t)}{\partial t}\frac{\partial t}{\partial T_i} \right)\right\vert_{t = k\delta-T_l}, \\
 \frac{\partial \boldsymbol{p}_j(t)}{\partial \boldsymbol{c}_j} &= \boldsymbol{\beta}(t),  \frac{\partial \boldsymbol{p}_j(t)}{\partial t} = \dot{\boldsymbol{p}}_j(t),  \frac{\partial t}{\partial T_i} = \begin{cases}
0, & i = j \\
-1, & i < j.
\end{cases}
\end{align}

The penalty cost functions are expressed as follows.



1) Static Obstacle Avoidance Penalty Cost $\mathcal{P}^{s}$: 
We maintain an Euclidean signed distance field (ESDF) map and query the distance and gradient from constraint points to obstacles, where the distance is denoted as $d(\overset{\circ}{\boldsymbol{p}}_k)$. The constraint points whose distance from obstacles is within a threshold will be penalized,
 and the gradients from the ESDF map will push them away from obstacles.


To enhance flight safety, an adaptive threshold $d^k_a$ considering FRS propagation is adopted. Define $d_q^k$ as the long semi-major axis of the following error bound ellipse $\boldsymbol{Q}^k$:
\begin{equation}
    \boldsymbol{Q}^k = \boldsymbol{Q}_{\mathrm{dist}}^k\oplus\boldsymbol{Q}_{\mathrm{ego}},
\end{equation}
where $\boldsymbol{Q}_{\mathrm{ego}}$ denotes the ellipsoid of the quadrotor's outer envelope. 
The disturbance estimation obtained from the observer is taken into account during the calculation of state variables, which is derived from the trajectory at the constraint points. The state variable is then involved in the update of $\boldsymbol{Q}_{\mathrm{dist}}^k$ in \eqref{eqn:qdist}. With $d_q^k$ being computed, the threshold is given by $d_a^k = d_q^k + d_s$, where $d_s$ denotes the static obstacle clearance.

Finally, the penalty of the $k$\textsuperscript{th} constraint point 
is formulated as
\begin{equation}
  P^s(\overset{\circ}{\boldsymbol{p}}_k) = \max \left\{ \psi^s(\overset{\circ}{\boldsymbol{p}}_k), 0 \right\}^3, 
\end{equation}
where $\psi^s$ is expressed as
\begin{equation}\label{psi^s}
 \psi^s(\overset{\circ}{\boldsymbol{p}}_k) = d^k_{a} - d(\overset{\circ}{\boldsymbol{p}}_k).
\end{equation}

2) Dynamic Obstacle Avoidance Penalty Cost $\mathcal{P}^{d}$: We adopt a constant-velocity model for dynamic obstacle motion prediction. Given a finite prediction horizon $\tau$, the predicted trajectory $\boldsymbol{p}_{\mathrm{pre}}(t)$ of the $\mu$\textsuperscript{th} moving obstacle at the world time $t_\mu$ can be represented as
\begin{equation}\label{pred_traj_dyn_obs}
    \boldsymbol{p}_{\mathrm{pre}}^{\mu}(t) = \boldsymbol{p}_{o}^{\mu} + \dot{\boldsymbol{p}}_{o}^{\mu}(t-t_\mu), \quad{t\in[t_\mu, t_\mu+\tau]},
\end{equation}
where $\boldsymbol{p}_o^{\mu}$ and $\dot{\boldsymbol{p}}_o^{\mu}$ are the position and velocity of the $\mu$\textsuperscript{th} obstacle acquired at $t_\mu$.

\begin{figure}[htbp]
\centering
\includegraphics[width=0.48\textwidth]{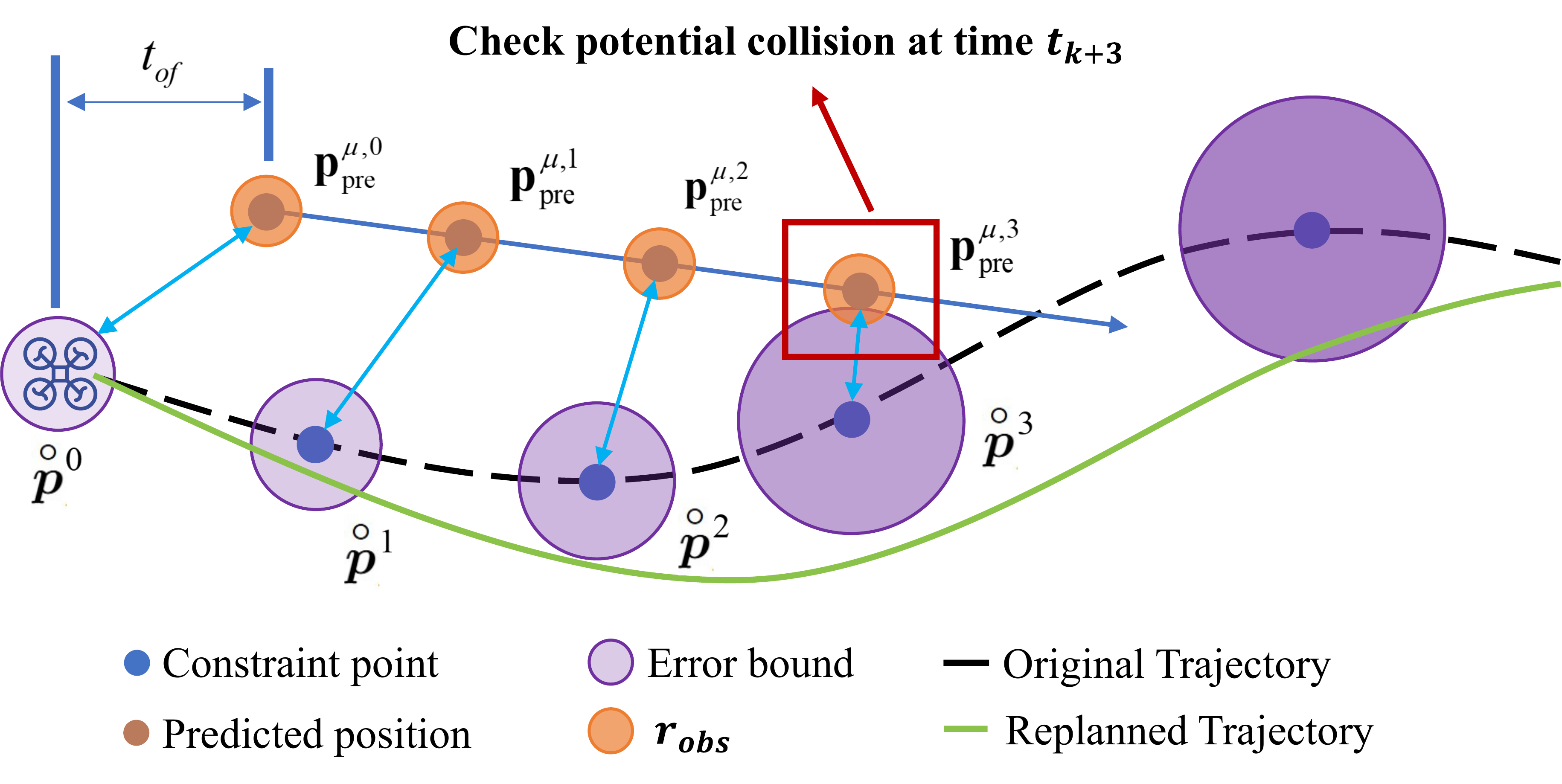}
\caption{Collision-check and replanning mechanism.}
\label{dyn_obs_check}
\end{figure}

Then, we build the dynamic obstacle constraint $\psi^d$ using \eqref{pred_traj_dyn_obs}:
\begin{equation}
    \psi^d\left(\overset{\circ}{\boldsymbol{p}}_k,\,\boldsymbol{p}^{\mu,k}_{\mathrm{pre}}\right) = (d_{r}^{k})^2 - \left\| \overset{\circ}{\boldsymbol{p}}_k- \boldsymbol{p}^{\mu,k}_{\mathrm{pre}} \right\|^2,
\end{equation}
where $d^k_{r} = d_q^k+d_c$ is the adaptive threshold with $d_c$ representing the dynamic obstacle clearance, and $\boldsymbol{p}^{\mu,k}_{\mathrm{pre}}\triangleq\boldsymbol{p}^\mu_{\mathrm{pre}}(k \delta+t_{of}+t_\mu)$ denotes the position of the $\mu$\textsuperscript{th} dynamic obstacle at the time corresponding to the 
$k$\textsuperscript{th} constraint point, with $t_{of}$ as an offset used to align the world time of the predicted trajectory with that of the quadrotor's trajectory. The dynamic obstacle avoidance mechanism is illustrated in Fig. \ref{dyn_obs_check}.
Since the time stamp of any constraint point is constant, $\boldsymbol{p}^{\mu,k}_{\mathrm{pre}}$ will not generate a gradient with respect to $\boldsymbol{T}$ \cite{quan2023robust}.
The penalty term for dynamic obstacle avoidance can be expressed as
\begin{equation}
\mathcal{P}^{d}\bigl(\overset{\circ}{\boldsymbol{p}}_k\bigr)
= \sum_\mu \max\Bigl\{\psi^{d}\left(\overset{\circ}{\boldsymbol{p}}_k,\,\boldsymbol{p}^{\mu,k}_{\mathrm{pre}}\right),\,0\Bigr\}^3.
\end{equation}

Furthermore, to address dynamic obstacles effectively, we employ a re-planning mechanism triggered by collision checking.
Firstly, we use the planning trajectory with error bound spheres and the predicted position of dynamic obstacles to check the relative distance between them at the same time stamp.
When the relative distance is less than the trigger threshold, the trajectory replan is activated.

3) Dynamic Feasibility Penalty Cost $\mathcal{P}^f$: We limit the maximum velocity and acceleration. The penalty is given by
\begin{subequations}
\begin{align}
\mathcal{P}^{f}(\overset{\circ}{\boldsymbol{p}}_k)&=\mathcal{P}^{f,v}(\overset{\circ}{\boldsymbol{p}}_k)+\mathcal{P}^{f,a}(\overset{\circ}{\boldsymbol{p}}_k), \\
\mathcal{P}^{f,v}(\overset{\circ}{\boldsymbol{p}}_k)&=\max\{\parallel{\overset{\circ}{\boldsymbol{p}}}{^\prime_k}\parallel^2-v_m^2,0\}^3, \\
\mathcal{P}^{f,a}(\overset{\circ}{\boldsymbol{p}}_k)&=\max\{\parallel\overset{\circ}{{\boldsymbol{p}}}{^{\prime\prime}_k}\parallel^2-a_m^2,0\}^3,
\end{align}
\end{subequations}
where $v_m$ and $a_m$ are the maximum velocity and acceleration.

\section{NMPC With Disturbance Compensation}
To ensure that the quadrotor can accurately track the planning trajectory, we utilize a generalized proportional integral observer \cite{chen2015disturbance} to compensate for the prediction model of NMPC:
\begin{equation}
  \begin{cases}
\dot{\hat{\boldsymbol{v}}} = \frac{1}{m} \left(\boldsymbol{R}(q)\,T_c - \boldsymbol{F}_{\text{drag}} + \hat{\boldsymbol{z}}_1\right) - \boldsymbol{g}  +\boldsymbol{G}_1\bigl(\tilde{\boldsymbol{v}} - \hat{\boldsymbol{v}}\bigr) , \\[6pt]
\dot{\hat{\boldsymbol{z}}}_{1} = \hat{\boldsymbol{z}}_{2} + \boldsymbol{G}_2\bigl(\tilde{\boldsymbol{v}} - \hat{\boldsymbol{v}}\bigr), \\[6pt]
\dot{\hat{\boldsymbol{z}}}_{2} = \boldsymbol{G}_3\bigl(\tilde{\boldsymbol{v}} - \hat{\boldsymbol{v}}\bigr),
\end{cases}
\end{equation}
where $\boldsymbol{G}_1, \boldsymbol{G}_2, \boldsymbol{G}_3 \in \mathbb{R}^{3\times 3}$ are the observer gains, $\boldsymbol{z}_1=\boldsymbol{F}_{\text{dist}}$,  
$\boldsymbol{z}_2=\dot{\boldsymbol{F}}_{\text{dist}}$, $\tilde{\boldsymbol{v}}$ denotes the velocity measurement, and $\hat{\boldsymbol{v}}$, $\hat{\boldsymbol{z}}_1$, $\hat{\boldsymbol{z}}_2$ denote the estimation of $\boldsymbol{v}$, $\boldsymbol{z}_1$, $\boldsymbol{z}_2$, respectively.

To facilitate NMPC design, we discretize the dynamic model of the quadrotor into $\boldsymbol{x}^{k+1} = f_d(\boldsymbol{x}^k,\boldsymbol{u}^k,\boldsymbol{F}_{\text{dist}})$ by forward Euler method with the time step $\Delta_t$. 
Then, we define the cost function including trajectory tracking error $\bigl\|\boldsymbol{x}^k - \boldsymbol{x}_{\mathrm{ref}}^k\bigr\|$ and control effort $\bigl\|\boldsymbol{u}^k-\boldsymbol{u}^k_{ref}\bigr\|$:
\begin{subequations}
\begin{align}
    J_x^k
& = \frac{1}{2}\bigl\|\boldsymbol{x}^k - \boldsymbol{x}_{\mathrm{ref}}^k\bigr\|_{\boldsymbol{G}_p},\\
J_x^N
& = \frac{1}{2}\bigl\|\boldsymbol{x}^N - \boldsymbol{x}_{\mathrm{ref}}^N\bigr\|_{\boldsymbol{G}_N}, \\
J_u^k
& = \frac{1}{2}\bigl\|\boldsymbol{u}^k-\boldsymbol{u}^k_{\mathrm{ref}}\bigr\|_{\boldsymbol{G}_u},
\end{align}
\end{subequations}
where $\boldsymbol{x}_{\mathrm{ref}}^k$ and $\boldsymbol{u}_{\mathrm{ref}}^k$ are the reference state and control input derived from planning trajectory, 
$\boldsymbol{G}_p, \boldsymbol{G}_N, \boldsymbol{G}_u \in \mathbb{R}^{3\times 3}$ are the weighted matrices.

Next, we incorporate disturbance estimation into the prediction model of NMPC, and the nonlinear optimization problem of NMPC is expressed as follows.
\begin{subequations}
\begin{align}
\min_{\boldsymbol{x},\,\boldsymbol{u}} \quad 
 J^N_x 
&+ \sum_{k=0}^{N-1} \Bigl(J_x^k + J_u^k\Bigr),
\label{eq:cost} \\[6pt]
\text{s.t.} \quad 
 \boldsymbol{x}^{k+1} &= f_d\bigl(\boldsymbol{x}^k, \boldsymbol{u}^k, \hat{\boldsymbol{F}}_{\mathrm{dist}}\bigr), \\
\quad \boldsymbol{x}^0 &= x_0, \\
\boldsymbol{x}^k &\in\mathbb{X}, \boldsymbol{u}^k\in\mathbb{U}, \boldsymbol{x}^N\in\mathbb{X}^N,
\label{eq:dynamics}
\end{align}
\end{subequations}
where $x_0$ is the initial state, and $\mathbb{X}, \mathbb{U}, \mathbb{X}^N$ are the constraint sets of state, control input, and terminal state, respectively.


\section{Simulation and Experiment}


The motion planning and control framework proposed in this paper is implemented in C++. We use ACADO\cite{Houska2011a} to solve the NMPC problem. The average computation time is 3 ms, and the control frequency is 100 Hz. For both simulation and real flight, an ESDF of the voxel grid map is maintained to get distance and gradient to obstacles for optimization. 

\subsection{Simulation Tests}
\begin{figure}[H]
\centering  
\subfigure[NMPC w/ the disturbance observer.]{
\label{Fig.sub.1}
\includegraphics[width=0.45\textwidth]{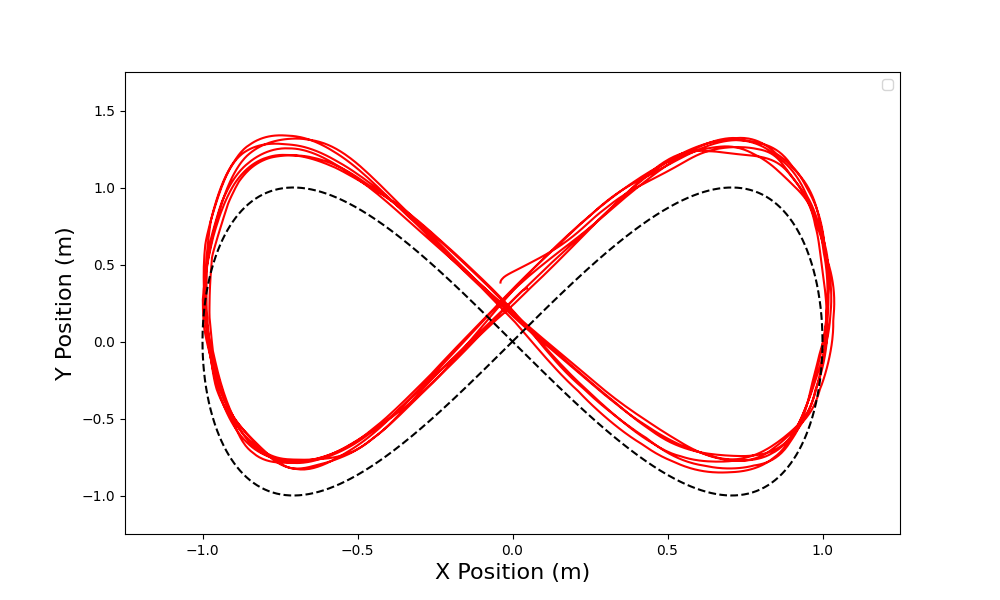}}
\subfigure[NMPC w/o the disturbance observer.]{
\label{Fig.sub.2}
\includegraphics[width=0.45\textwidth]{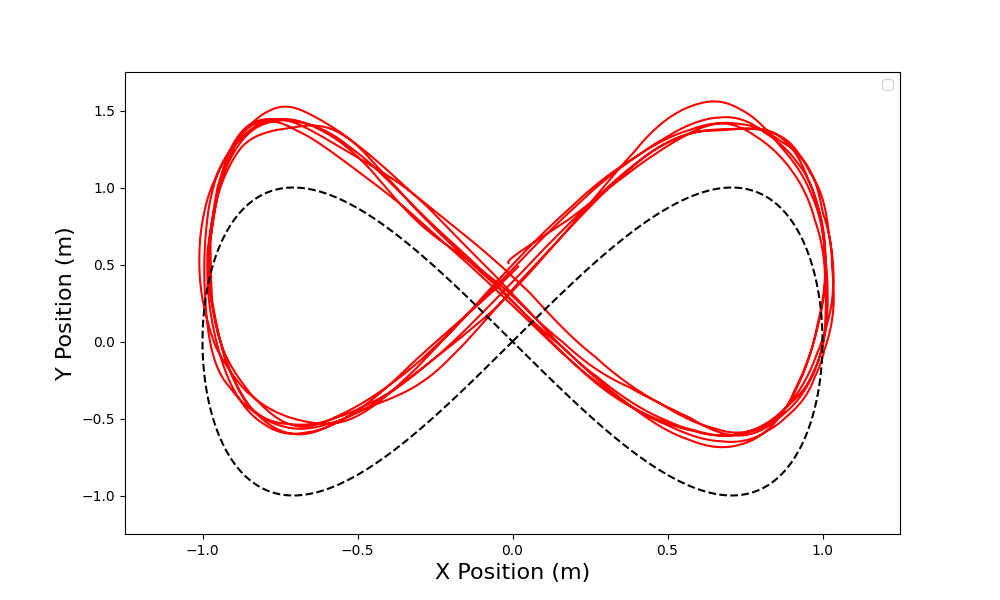}}
\caption{Quadrotor trajectory in X-Y plane under wind with average speed of $5m/s$.}
\label{tracking_comparison}
\end{figure}

\begin{figure}[htbp]
\centering
\includegraphics[width=0.42\textwidth]{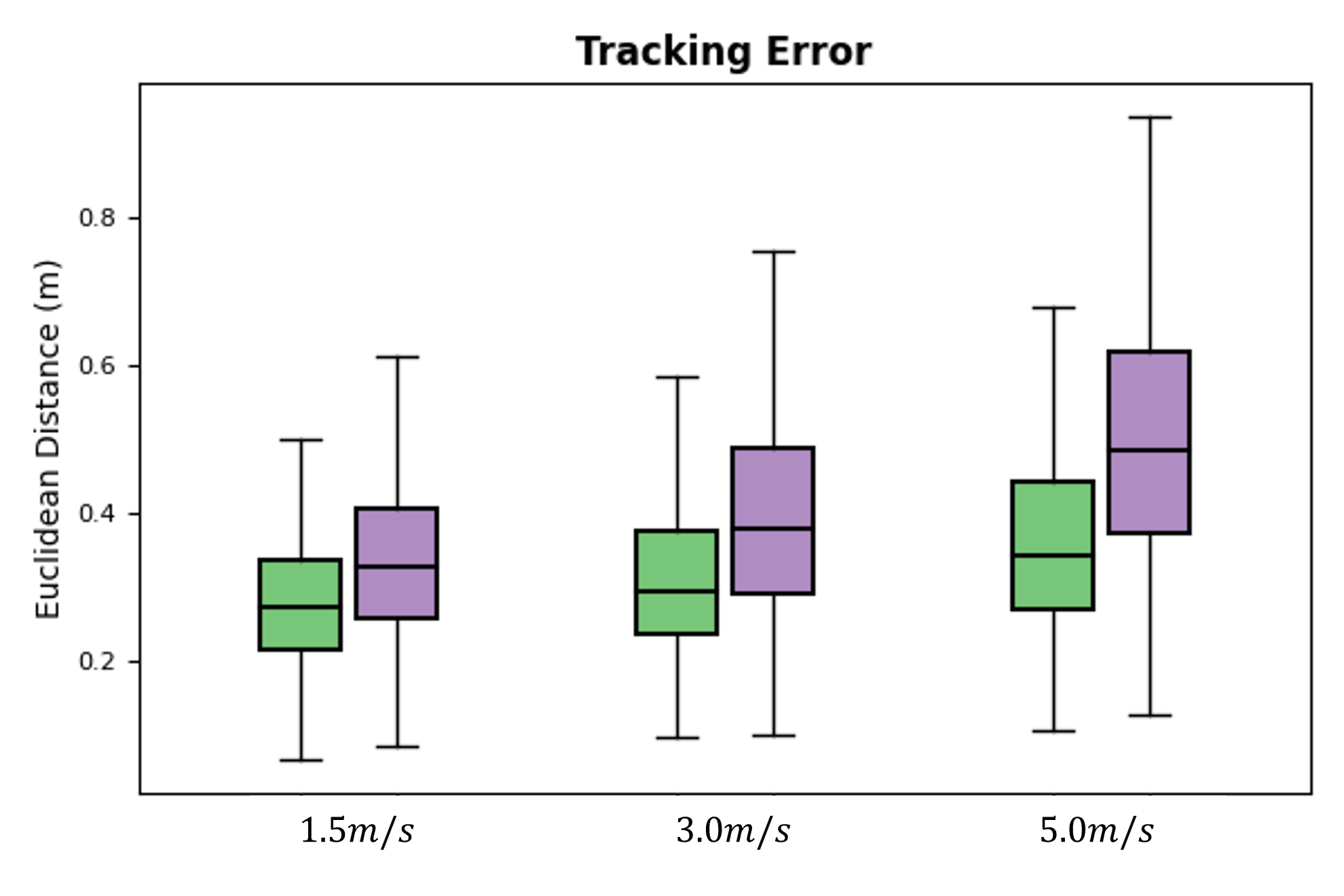}
\caption{The tracking error of NMPC w/ and w/o disturbance observer.}
\label{test_data}
\end{figure}

1) NMPC Ablation Experiment: We first evaluate NMPC with and without the disturbance observer on tracking the same reference eight trajectory in Fig. \ref{tracking_comparison} against the wind. The maximum velocity The average wind speed ranges from $1.5m/s$ to $5m/s$, and the variance is $1m^2/s^2$. 
The performance is measured as Min/Max/Avg in Fig. \ref{test_data}. The green box is NMPC with the disturbance observer, and the purple box is the baseline NMPC. These illustrate that NMPC with the disturbance observer has better tracking performance due to further disturbance compensation. On average, the use of the disturbance observer can reduce nearly $22.9\%$ of the tracking error. 

\begin{figure}[htbp]
\centering
\includegraphics[width=0.35\textwidth]{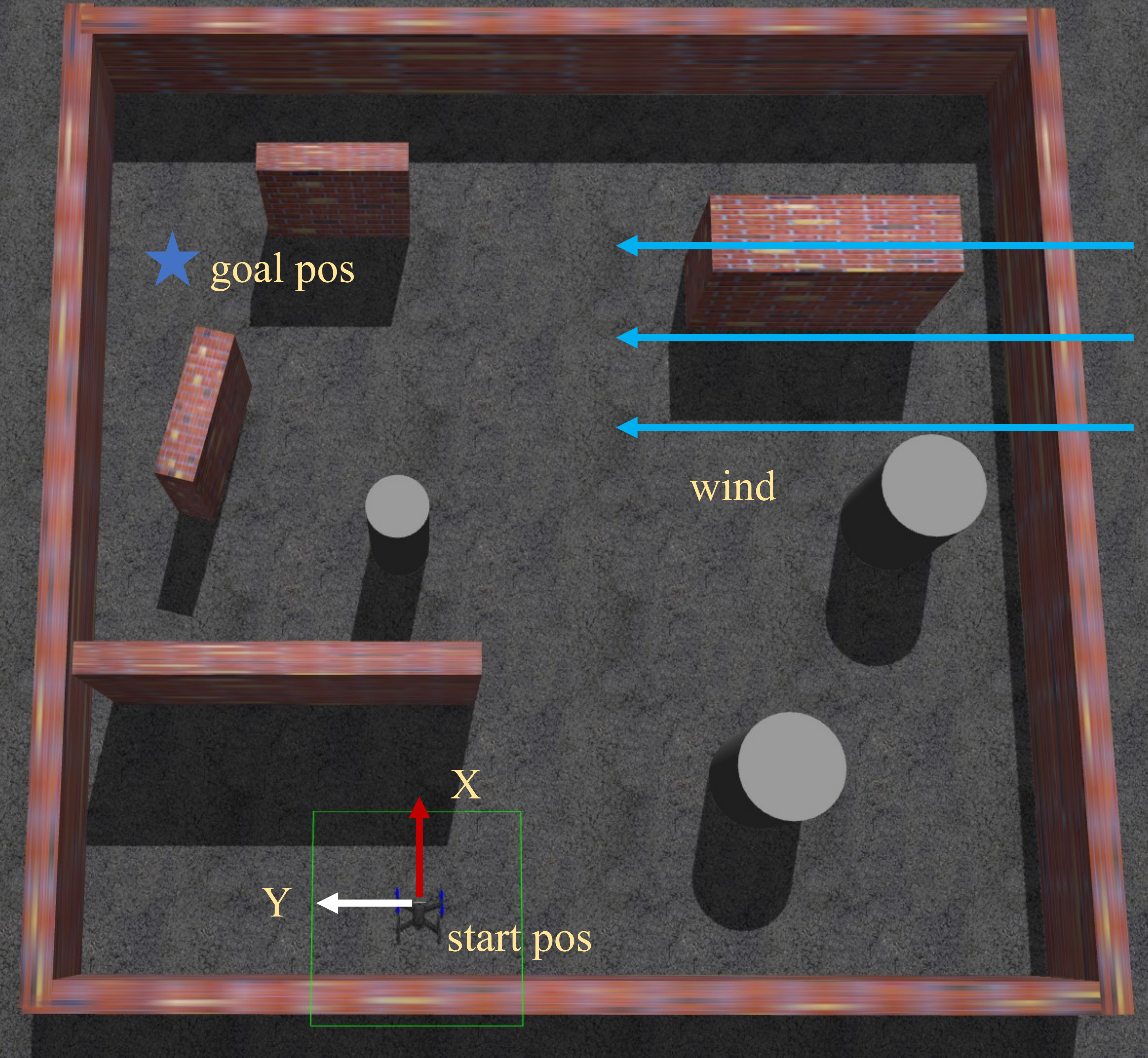}
\caption{The simulation environment in Gazebo.}
\label{simu_env}
\end{figure}

2) Comparison of the planning methods.
We compare our method with a state-of-the-art local planner, EGO-Planner V2 \cite{zhou2022swarm}, which is also based on MINCO. 
We use Gazebo as the simulator and PX4 as the low-level controller to achieve relatively realistic simulations. The simulation environment is shown in Fig. \ref{simu_env}. As a wind zone fully covers the whole map, the quadrotor is under continuous disturbances along the positive y-axis with changing wind speed. 30 trials are conducted per condition.

\begin{table}[h]  
	\centering     
    \renewcommand{\arraystretch}{1.1}
	\caption{Comparison of the Planning Method}  
	\label{comparison}        
	\begin{tabular}{ c c c c  }  
		\toprule[1.5pt]   
		
		\multicolumn{2}{c}{\textbf{Wind Field Parameters}}  &  &  \\ \cline{1-2}

               Avg. Speed ($m/s$) & $\sigma^2$ ($m^2/s^2$)    & \multirow{1}{*}[1.5ex]{\centering \textbf{Method}}    &  \multirow{1}{*}[1.5ex]{\centering \textbf{Succ. Rate}}   \\ 
		
		\midrule[1pt]     
		
		\multirow{2}{*}{$3.5$}  & \multirow{6}{*}{$1$}  & Ours & 1.00 \\ 
		                        &	               & Zhou\cite{zhou2022swarm} & 1.00 \\ 

            \multirow{2}{*}{$5$}  &               & Ours & \textbf{1.00} \\ 
		                        &	               & Zhou\cite{zhou2022swarm} & 0.27 \\ 
                                
            \multirow{2}{*}{$6.5$}  &              & Ours & \textbf{0.86} \\ 
		                        &	               & Zhou\cite{zhou2022swarm} & 0.00 \\

		\bottomrule[1.5pt]   
	\end{tabular} 
\end{table}

During the simulation tests, the NMPC controller integrated with a disturbance observer was employed for both planning methods. 
The success rate, defined as the proportion of flights conducted without any collisions, demonstrates distinct performance characteristics under varying wind conditions. 
As shown in Table \ref{comparison}, Zhou’s method \cite{zhou2022swarm} exhibits a precipitous decline in success rate with increasing wind speed. Particularly noteworthy is that at the critical wind speed of $6.5\text{ }m/s$, the proposed method maintains a success rate of 0.86, whereas the comparative baseline fails catastrophically across all test cases.

\begin{figure}[htbp]
\centering
\includegraphics[width=0.45\textwidth]{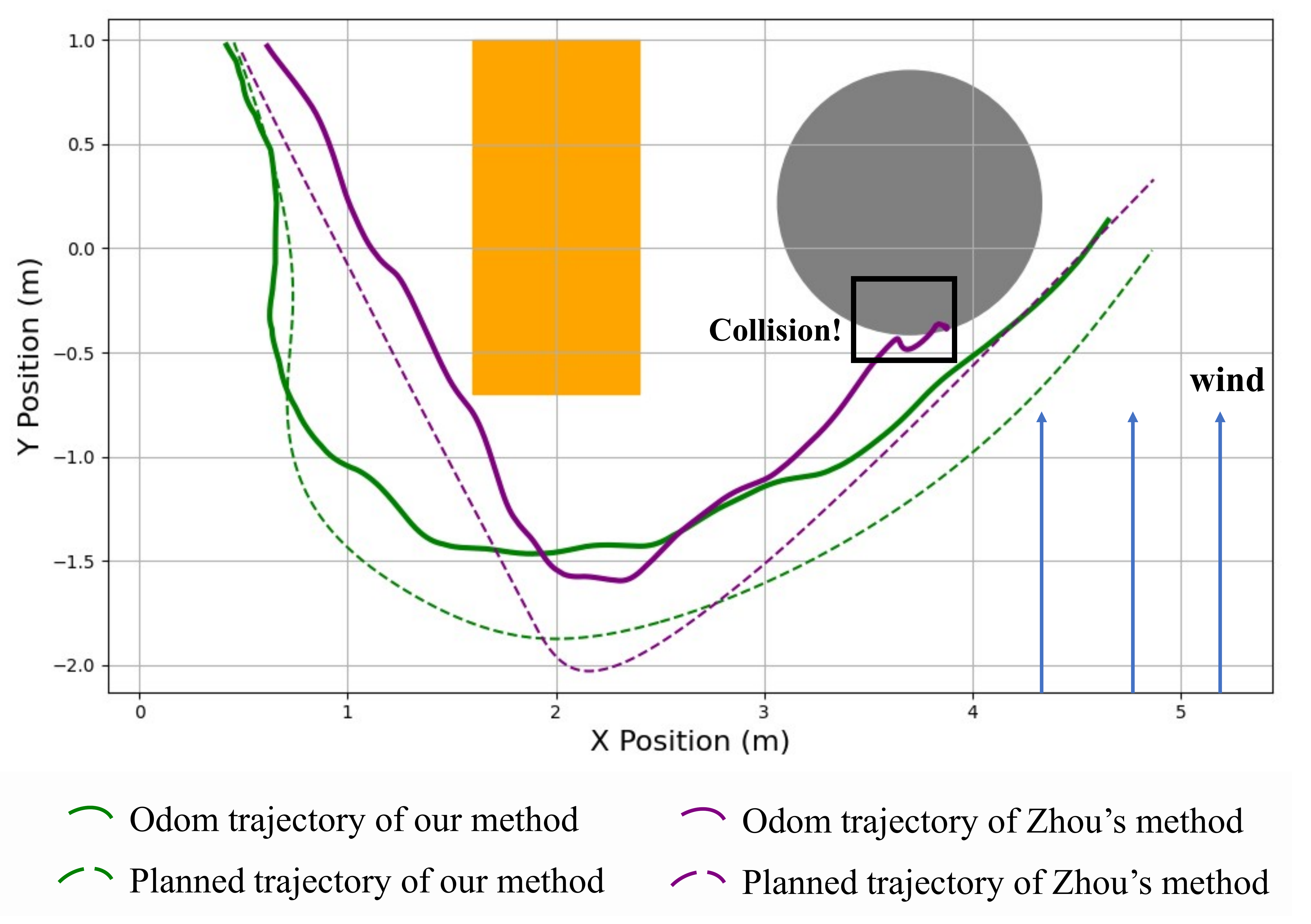}
\caption{The trajectory of one simulation test under a wind speed of $6.5\text{ }m/s$. The orange rectangle and the gray circle are the obstacles after inflation by the radius of the quadrotor. The wind field covers the entire map along the positive y-axis.}
\label{collision}
\end{figure}

Mild disturbance ($3.5\text{ }m/s$ wind) can be compensated by the controller. However, as the wind speed is up to $6.5\text{ }m/s$, the disturbance mitigation ability of the controller is reduced, and safety can not be ensured in Zhou's method \cite{zhou2022swarm}. Furthermore, even if the wind disturbance is estimated by the disturbance observer and compensated for within the NMPC controller, the quadrotor still cannot fly safely due to the lack of proactive consideration in the planning phase. In contrast, our planner considers the wind disturbance FRS, allowing the quadrotor to execute a safer trajectory, as demonstrated in Fig. \ref{collision}.

\subsection{Real-World Flight}
\begin{figure}[htbp]
\centering
\includegraphics[width=0.35\textwidth]{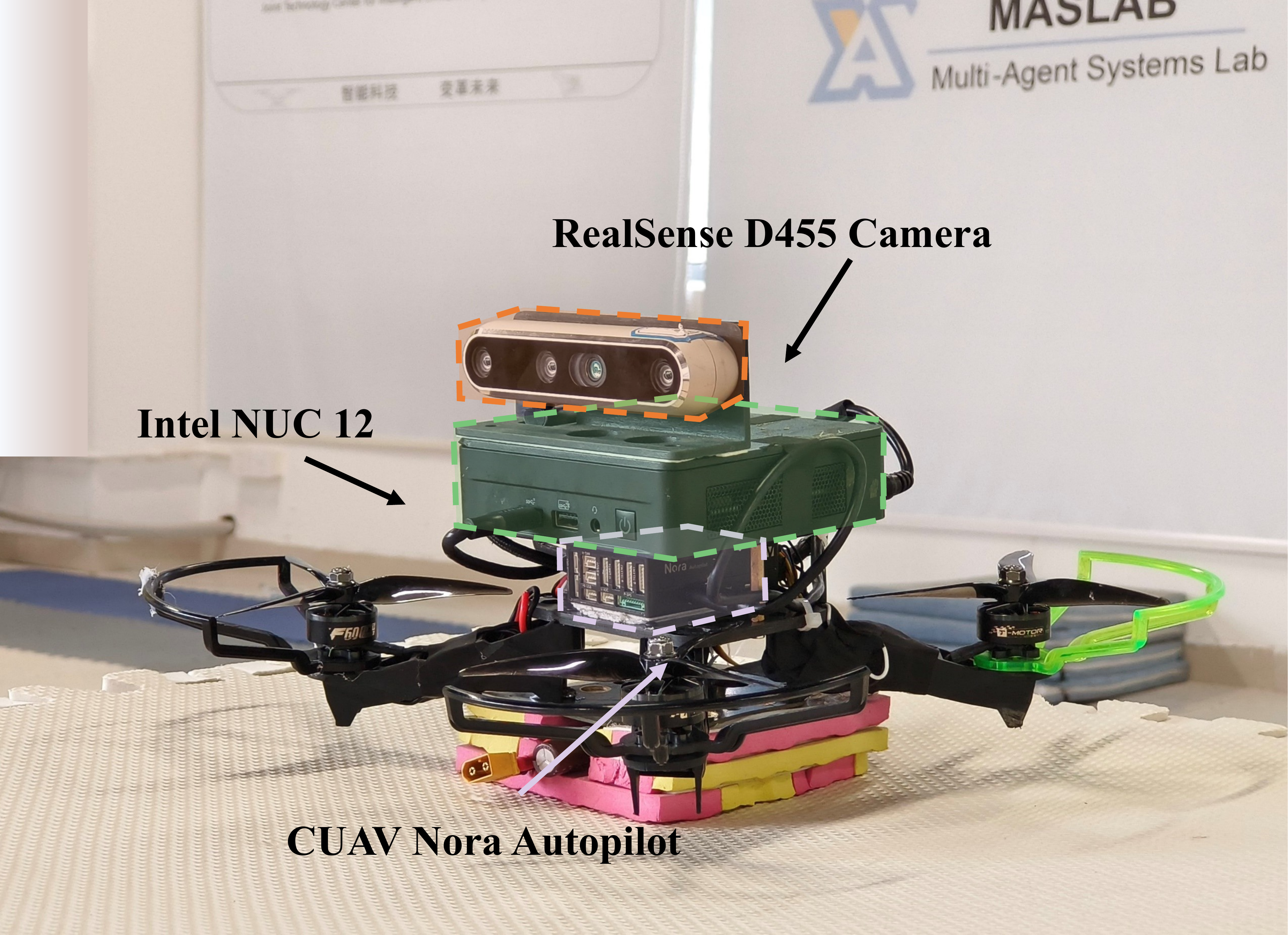}
\caption{Quadrotor platform for real flight.}
\label{drone250}
\end{figure}

The quadrotor platform that we use is shown in Fig. \ref{drone250}. All the perception, planning, and control algorithms are run on an onboard Intel NUC 12 computer with CPU i7-1260P and 16GB RAM. Localization and mapping are from VINS-Fusion \cite{qin2019general}. The velocity of the dynamic obstacle is measured by the NOKOV motion capture system. 

We present an indoor flight experiment with wind and a dynamic obstacle to validate our framework, and the snapshot is shown in Fig. \ref{flight_snap} (a). The dynamic obstacle moves back and forth with an average speed of $0.8\text{ }m/s$. A fan near the area that the drone will pass is to provide wind disturbance, and the wind speed ranges from $3.2\text{ }m/s$ to $4.5\text{ }m/s$. For safety considerations, the maximum speed and acceleration of planning are $2\text{ }m/s$ and $6\text{ }m/s^2$, respectively. 

\begin{figure}[H]
\centering  
\subfigure[The scene of the real-world flight.]{
\label{Fig.sub.1}
\includegraphics[width=0.48\textwidth]{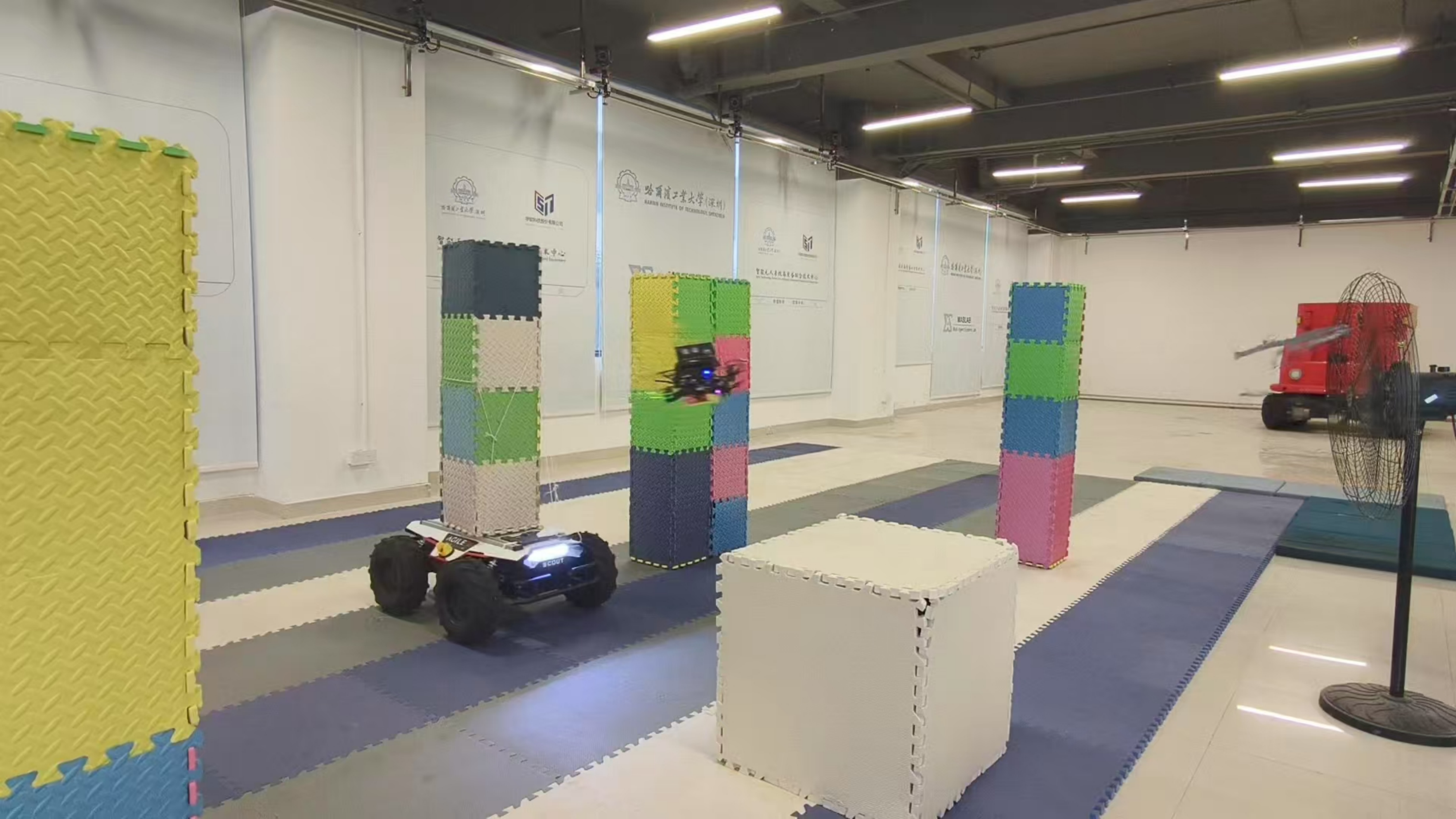}}
\subfigure[Visualization of the real-world flight.]{
\label{Fig.sub.2}
\includegraphics[width=0.48\textwidth]{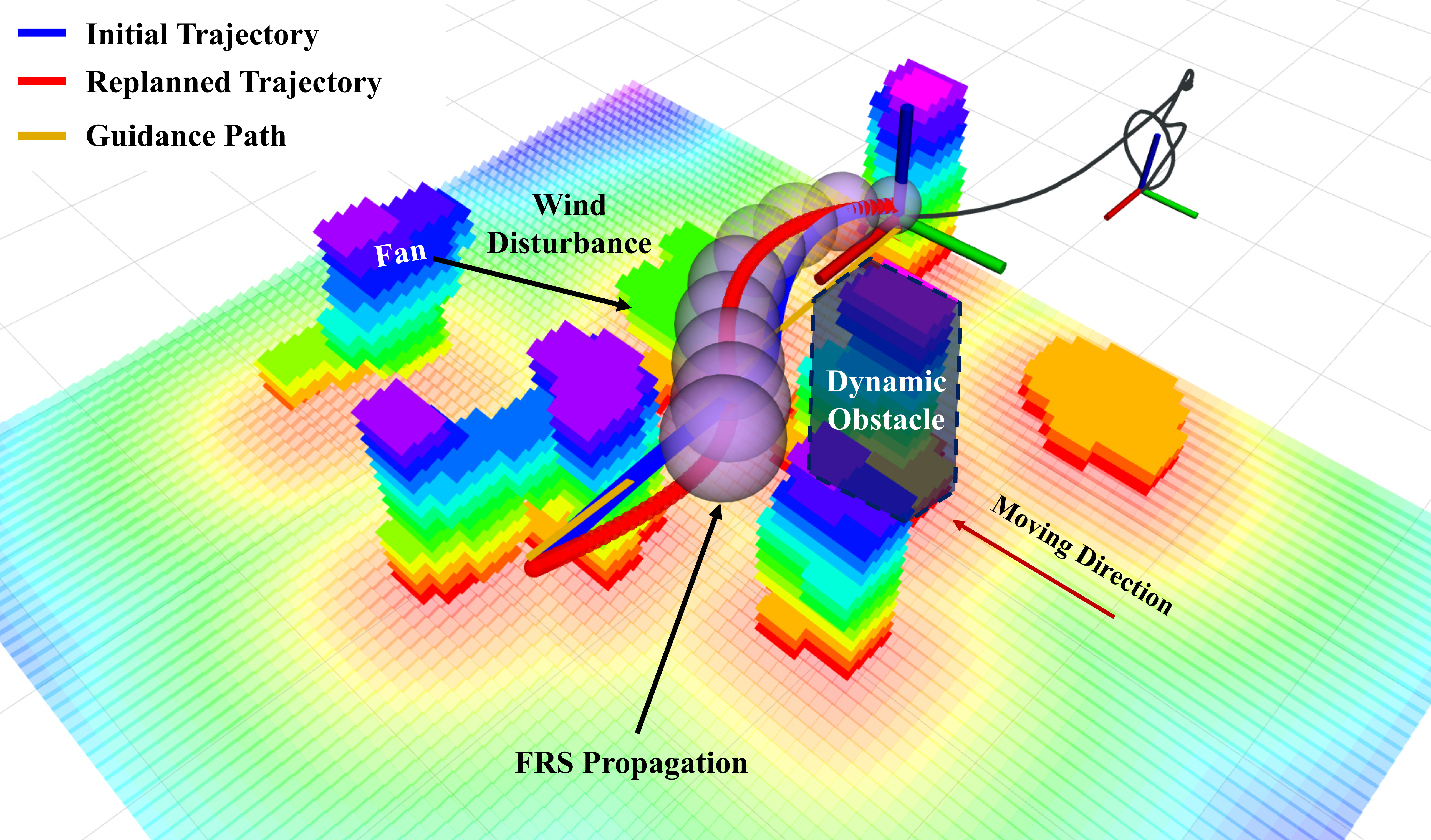}}
\caption{Trajectory replanning and tracking in consideration of dynamic obstacle and wind disturbance. The dynamic obstacle is simulated using an AGILEX UGV with a column-shaped obstacle mounted on it, and the wind disturbance is generated by a fan.}
\label{real}
\end{figure}

In Fig. \ref{real}, we illustrate the replanning process during the real flight. As the dynamic obstacle moves towards the quadrotor, our framework detects the potential collision. Therefore, the replanning is triggered for a new trajectory (the red line), and the initial one (the blue line) will be discarded. 
Even under wind disturbances, the new trajectory is pushed away from the dynamic obstacle.


Subsequently, a safe and smooth trajectory is generated as a reference to the disturbance-aware tracking controller. The tracking error during the flight is in Tab. \ref{tracking_error}.  

\begin{table}[h]
\caption{Trajectory Tracking Error}
\label{tracking_error}
\renewcommand{\arraystretch}{1.25}
\begin{center}
\begin{tabular}{ccccc}
\toprule[1.5pt]
Avg ($m$) & Min ($m$) & Max ($m$) & RMSE ($m$)\\
\hline
 0.1378 & 0.0019 & 0.1902 & 0.0211\\
\bottomrule[1.5pt]
\end{tabular}
\end{center}
\end{table}

\section{Conclusions}
In this work, we present a systematic safety-enhanced trajectory planning and control framework for quadrotors under the dynamic obstacles environment with wind disturbances. By handling wind disturbances in both planning and control using the information obtained from the disturbance observer, the proposed method ensures collision avoidance for autonomous flight. 
Simulation tests and real-world flights have validated the robustness to wind disturbances and 
 the dynamic obstacles.
Our future work will focus on extending this method to a large-scale distributed quadrotor swarm system.

\addtolength{\textheight}{-5.5cm}   








\bibliographystyle{IEEEtran}
\bibliography{IEEEabrv,ref}

\end{document}